\documentclass[12pt]{article}
\usepackage{amsmath,amssymb,amsthm,color}
\usepackage{graphicx}
\usepackage{cite}
\usepackage{caption}
\usepackage{epsfig}
\usepackage{epstopdf} 
\renewcommand{\baselinestretch}{2}
\begin{document}
%
\title{$\pi$-bonding-dominated energy gaps in graphene oxides\\}
\author{
\small Ngoc Thanh Thuy Tran$^{a}$, Shih-Yang Lin$^{a}$, Olga E.Glukhova$^{b}$, Ming-Fa Lin$^{a,*}$ $$\\
\small $^a$Department of Physics, National Cheng Kung University, Tainan 701, Taiwan \\
\small $^b$Department of Physics, Saratov State University, Saratov 410012, Russia\\
 }
\renewcommand{\baselinestretch}{1}
\maketitle

\renewcommand{\baselinestretch}{1.4}
\begin{abstract}

Chemical bondings of graphene oxides with oxygen concentration from 1\% to 50\% are investigated using first-principle calculations. Energy gaps are mainly determined by the competition of orbital hybridizations in C-C, O-O, and C-O bonds. They are very sensitive to the changes in oxygen concentration and distributions. There exists five types of $\pi$ bondings during the variation from the full to vanishing adsorptions, namely the complete termination, the partial suppression, the 1D bonding, the deformed planar bonding, and the well-behaved one. They can account for the finite and gapless characteristics, corresponding to the O-concentrations of $>$25\% and $<$3\%, respectively. The feature-rich chemical bondings dominate band structures and density of states, leading to diverse electronic properties. 
\vskip 1.0 truecm
\par\noindent

\vskip1.0 truecm

\par\noindent  * Corresponding authors. {~ Tel:~ +886-6-2757575-65272 (M.F. Lin)}\\~{{\it E-mail address}: mflin@mail.ncku.edu.tw (M.F. Lin)}
\end{abstract}
\pagebreak
\renewcommand{\baselinestretch}{2}
\newpage

{\bf 1. Introduction}
\vskip 0.3 truecm

Graphene has been a remarkable material with potential applications in many areas such as transistors \cite{C5NR03294F}, sensors \cite{C5RA00871A}, and photovoltaic cells \cite{C4TA01047G}. However, it is limited by the zero-gap characteristics which needs to be significantly modified for further developments of graphene-based devices. Numerous methods and techniques have been developed to solve this problem, including some based on doping \cite{liu2011chemical,nakada2011migration}, external electric or magnetic fields \cite{tang2011electric}, and nanomesh \cite{C4NR04584J}. Understanding the electronic properties of the doped graphene is a main-stream topic in physics, chemistry and materials. Recently, graphene oxides (GO) has stirred much attention due to its tunable band gap and other interesting properties \cite{mkhoyan2009atomic,ito2008semiconducting}. GO is considered as a promising material for fabrication of energy storage and environmental protector, including supercapacitors \cite{Xue2015305,chen2011high}, lithium batteries \cite{li2015graphene}, memristor devices \cite{Porro2015383}, sensors \cite{veerapandian2012synthesis,robinson2008reduced}, and water purification \cite{li2015graphene}. This paper focuses to investigate the dependence of the energy gap on O-concentration and the responsible mechanisms associated with the chemical bondings in C-C, C-O and C-C bonds.

Brodie, Staudenmaier, and Hummers are the three major methods to produce GO from graphite in which the third method is widely used because of its shorter reaction time and no ClO$_2$ emission \cite{poh2012graphenes,bai2011functional}. Recently, GO has also been manufactured using Tang-Lau method ($"$bottom-up$"$ method) since its process is simpler and more environmentally friendly compared to traditionally $“$top-down$”$ methods \cite{tang2012bottom}. Different O-concentrations can be synthesized by controlling the amount of oxidant compound (KMnO$_4$) \cite{C5RA02099A}, or the oxidation time \cite{C4RA02873B}. The concentrations and distributions are further examined using nuclear magnetic resonance \cite{cai2008synthesis}. In addition, X-ray photoelectron spectroscopy is a powerful instrument, which can provide information on how many atoms are doped into graphene and the types of the adatoms \cite{gao2009new}. On the theoretical side, the previous studies show that the zero gap of graphene is significantly opened by oxygen adsorption \cite{lian2013big,sutar2012electronic,ito2008semiconducting}. However, a systematic study about the dependence of energy gap on the wide range of O-concentration has not been investigated. The full information about the geometric structure and electronic properties of GO with O-concentration from 1\% to 50\% need to be explored in details. This paper seeks to comprehend the critical mechanisms of the O-dependent energy gaps, being related to the chemical bondings.

The effect of oxygen, with different geometric symmetries and concentrations, on the electronic properties of GO are investigated using first-principle calculations. The results demonstrate that the chemical bondings, in the C-C, C-O, and O-O bonds, are very sensitive to changes in O-coverage. The complicated relations among these bondings lead to diverse electronic properties, including the destruction or distortion of Dirac cone, opening of band gap, and energy dispersions related to O-O and C-O bonds. Such features are reflected in the orbital-projected DOS, such as the absence and presence of $\pi$ band peak, the O-dominated special structures, and the (C,O)-dominated prominent peaks. The distorted Dirac-cone and the opening of band gap will be explored through the 3D band structures and band decomposed charge density analysis to verify their relation with the $\pi$ bondings. The above-mentioned results can be verified by experimental measurements such as angle resolved photoemission spectroscopy (ARPES) \cite{ohta2006controlling}, and scanning tunneling spectroscopy (STS) \cite{stolyarova2007high}.

\vskip 0.6 truecm
\par\noindent
{\bf 2. Methods }
\vskip 0.3 truecm

The first-principle calculations on GO are performed based on density functional theory using the Vienna Ab initio Simulation Package \cite{kresse1996efficient,kresse1999ultrasoft}. The electron-ion interactions are taken into account by the projector augmented wave method \cite{blochl1994projector}, whereas the electron-electron interactions are evaluated using the exchange-correlation function under the generalized gradient approximation of Perdew-Burke-Ernzerhof \cite{perdew1996generalized}. A vacuum layer with a thickness of 14 $\mbox\AA$ is added in a direction perpendicular to the GO plane to avoid interactions between adjacent unit cells. The wave functions are expanded using a plane-wave basis set with a maximum kinetic energy of 500 eV. All atomic coordinates are relaxed until the total energy difference is less than 10$^{-5}$ eV, and the Hellmann-Feynman force on each atom is less than 0.01 eV/$\mbox\AA$. The k-point mesh is set as 100$\times$100$\times$1 in geometry optimization, 30$\times$30$\times$1 in band structure calculations, and 150$\times$150$\times$1 in the DOS calculations for the 2$\times$2 supercell. An equivalent k-point mesh is set for other supercells depending on their sizes. 

\vskip 0.6 truecm
\par\noindent
{\bf 3. Results and discussion}
\vskip 0.3 truecm

\begin{figure}[htb]
\centering\includegraphics[width=12cm]{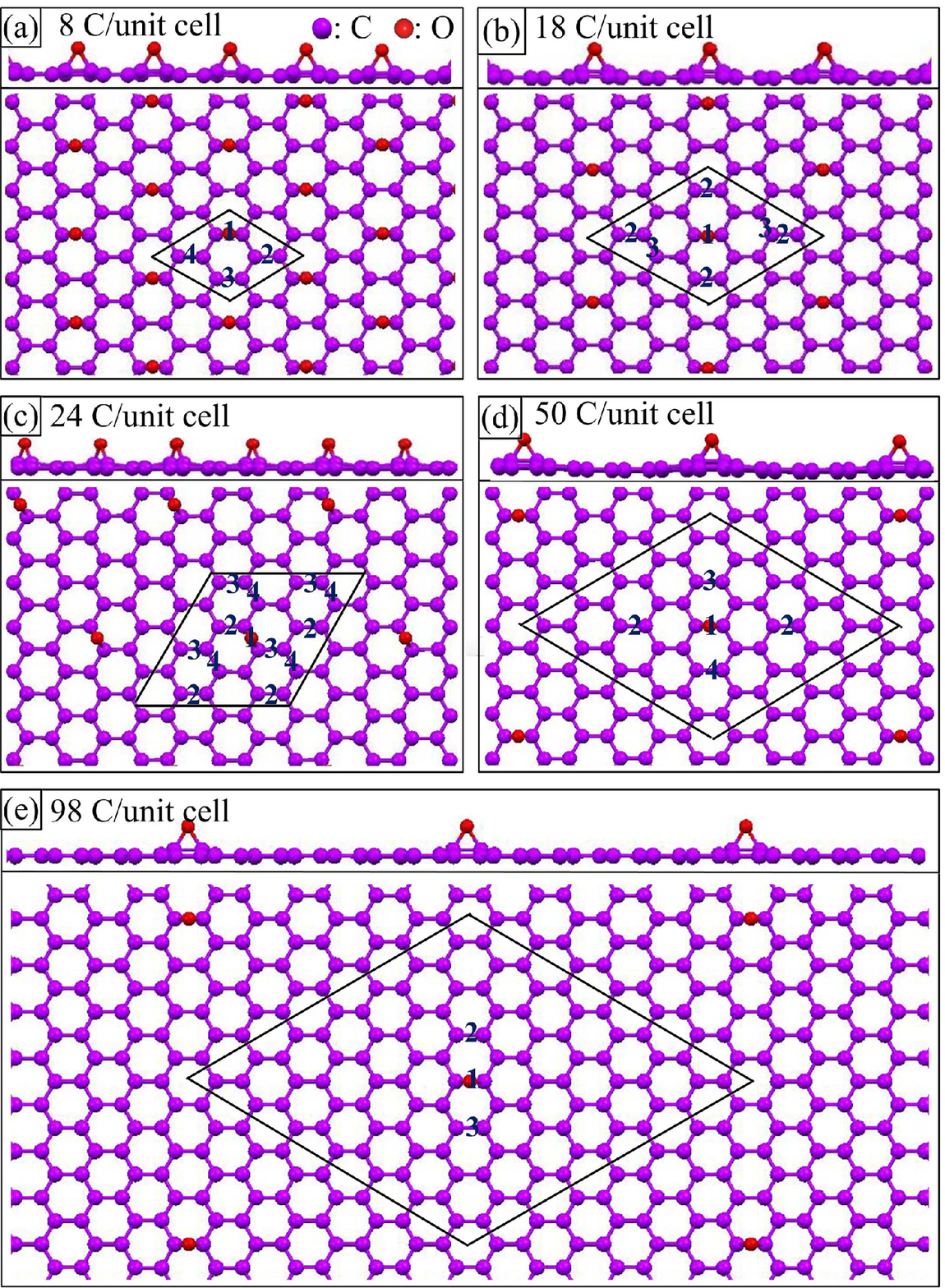}
\caption{Geometric structures of GO with various unit cells and concentrations: (a) 8 C atoms per cell, (b) 18 C atoms per cell, (c) 24 C atoms per cell, (d) 50 C atoms per cell, and (e) 98 C atoms per cell. Oxygen distributions are denoted by numbers, corresponding to different concentrations listed in table 1.}
\end{figure}

The atomic structures of oxygen adsorbed monolayer graphene are shown in Fig. 1. O atoms are at the top between two carbon atoms as so-called the bridge-site with O/C ratio from 1\% to 50\%. From the total ground state energies, the bridge-site is the most stable position compared to the hollow- and top-sites, which agrees with previous studies \cite{nakada2011migration}. After the self-consistent relaxations, all the GO systems, with various coverages, have stable structures in the presence of  a very slight buckling. The binding energy, characterizing the reduced energy due to the oxygen adsorption, is defined as $E_{b}$ = $(E_{sys}$ - $E_{gra}$ - n$E_{O}$)/n, where $E_{sys}$, $E_{gra}$, and $E_{O}$ are the total energies of the GO system, the graphene sheet, and the isolated O atoms, respectively. In general, the stabilities are enhanced by the higher O-concentration and distribution symmetry. It is worth to mention that the relative stability among GO configurations for each O-concentration is only slightly changed, $\approx$ 1-5\% difference. The main features of geometric structures, including the optimized C-C bond lengths, and C-O bond lengths strongly depend on the O-concentration, as shown in Table 1. The C-C bond lengths are expanded due to oxygen adsorption and gradually recover to that in pristine graphene (1.42 $\mbox\AA$) in the decrease of O-concentration, which agrees well with previous studies \cite{mkhoyan2009atomic,ito2008semiconducting}. However, the C-O bond lengths exhibit the opposite dependence. These results indicate that the stronger orbital hybridizations in C-O bonds significantly reduce their lengths and even weaken the sigma bondings in C-C bonds. The above-mentioned rich geometric structures lead to diverse electronic properties.

\begin{table}[htb]
 \caption{The calculated C-O bond lengths, C-C bond lengths, binding energy and energy gap ($E_g$) for various concentrations. The electronic structures of concentrations labeled with * will be illustrated in detail later.}
 \label{t1}
 \begin{center}
 \begin{tabular}{ c c c c c c c c }
 \hline
Figure&Atoms&Oxygen&C-O bond&C-C bond&Binding&$E_g$ \\
& &concentration&length (\AA)&length (\AA)&energy(eV)&(eV)\\
\hline
(a) & 1, 2, 3 ,4 & 50\%$^*$ & 1.43 & 1.53 &-4.4651 &3.54\\
(c) & 2, 3 & 33.3\%$^*$ & 1.45 & 1.48 &-3.7433 &2.53\\
(c) & 2, 4 & 33.3\% & 1.45 & 1.48 &-3.5751&1.28\\
(b) & 1, 2 & 27.8\% & 1.45 & 1.47 &-4.0788 &1.26\\
(a) & 1, 3 & 25\%$^*$ & 1.46 & 1.46 &-4.0093 &1.37\\
(a) & 1, 2 & 25\% & 1.46 & 1.46&-3.6509 & 0.63\\
(b) & 2 & 22.2\%$^*$ & 1.46 & 1.45 &-4.0989& 0\\
(b) & 1, 3 & 16.7\% & 1.46 & 1.44 &-3.9557& 0.36\\
(c) & 2 & 16.7\% & 1.46 & 1.44 &-3.8586& 0.09\\
(a) & 1 & 12.5\%$^*$ &1.46 & 1.44 &-3.9487& 0\\
(d) & 1, 2, 3, 4 & 10\%$^*$ & 1.46 & 1.44 &-3.9576& 0.21\\
(c) & 1 & 4.2\%$^*$ & 1.47 & 1.43&-3.9636 & 0.07\\
(e) & 1, 2, 3 & 3\%$^*$ & 1.47 & 1.42 &-3.9245& 0\\
(d) & 1 & 2\%$^*$ & 1.47 & 1.42 &-3.9187& 0\\
(e) & 1 & 1\%$^*$ & 1.47 & 1.42 &-3.8918& 0\\
\hline
 \end{tabular}
 \end{center}
 \end{table}
 
The two-dimensional energy bands along of GO high symmetric points alter dramatically with the variations in O-concentration, as shown in Fig. 2. The contributions of passivated C atoms and O atoms are represented by the red and blue circles, respectively, in which the dominance is proportional to the circle's radius. With O-concentration in the range of 25-50 \% (Figs. 2(a)-2(c)), the isotropic Dirac-done structure of pristine graphene is destroyed; furthermore,  the opening of large energy gaps is mainly determined by the O-dominated energy band nearest to $E_F$ (blue circles). That is, the $\pi$ bonding is replaced by the orbital hybridizations of O-O bonds.  As to the strong orbital hybridizations of passivated C atoms and O atoms (red and blue circles), their energy bands are located in -2 eV$\,\le\,E^v\le\,$-4 eV. The deeper $\sigma$-bands with $E^v\le\,$-3.5 eV are formed by the rather strong hybridization of (2$p_x$,2$p_y$,2s) orbitals in C-C bonds almost independent of O-adsorption. With the decrease of O-concentration, the $\pi$-band is gradually recovered, while the O-dominated bands become narrower and contribute at deeper energy, while the $\pi$ bands due to the neighboring bondings of parallel 2$p_z$ orbitals are gradually recovered (Figs. 2(d)-2(k)). Their competition leads to the diverse energy dispersions in the middle- and low-concentration GO systems, e.g., the distorted Dirac-cone structures with narrow or zero gaps (insets). 

\begin{figure}[htb]
\centering\includegraphics[width=9cm]{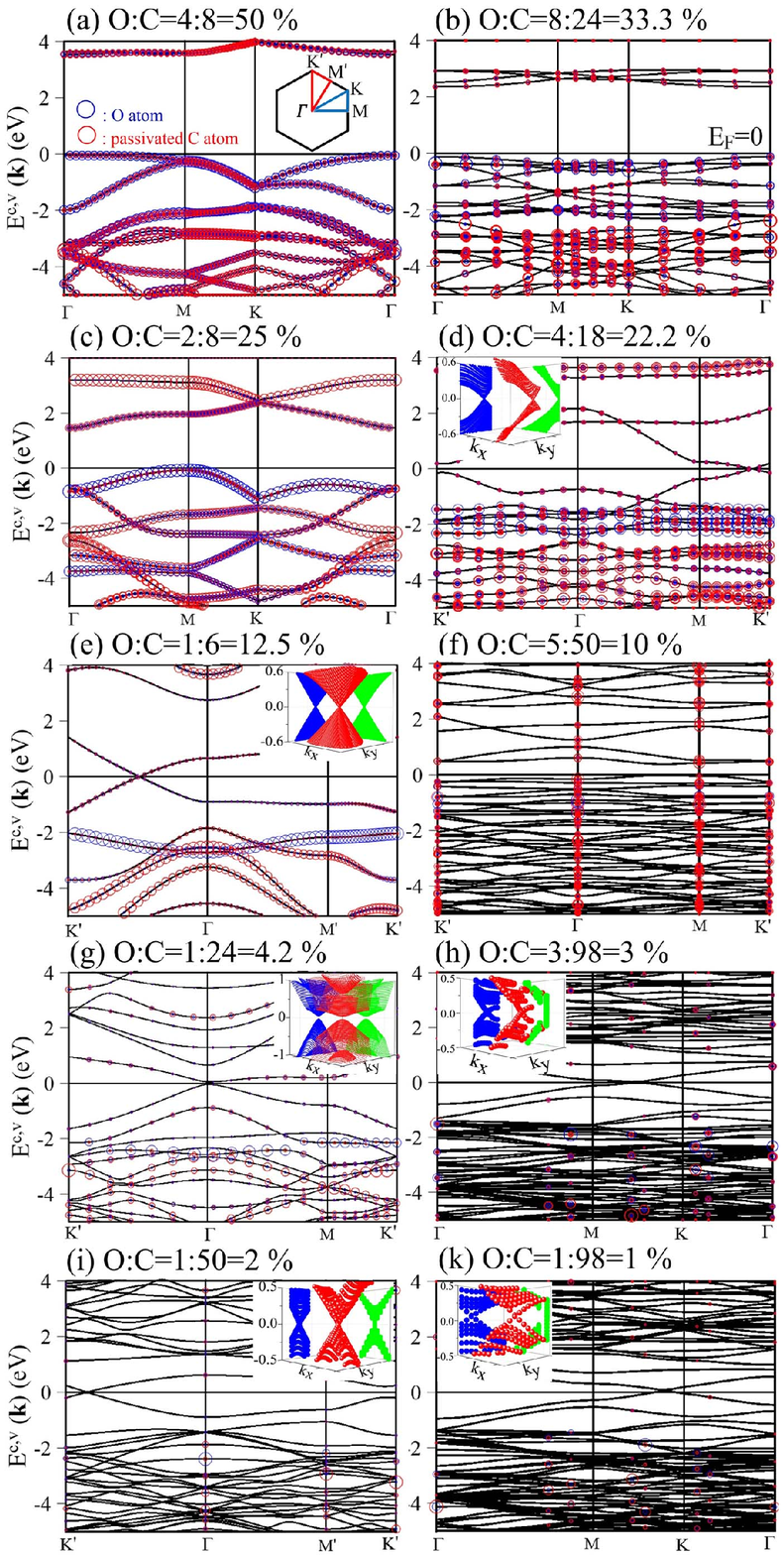}
\caption{Band structures of GO with various concentrations: (a) 50\%, (b) 33.3\%, (c) 25\%, (d) 22.2\%, (e) 12.5\%, (f) 10\%, (g) 4.2\%, (h) 3\%, (i) 2\%, and (k) 1\%. Superscripts $c$ and $v$ correspond to the conduction and valence bands, respectively. The red and blue circles, respectively, correspond to the contributions of passivated C atoms and O atoms, in which the dominance is proportional to the radius of circle. Also shown in the inset of (a) is the first Brillouin zone.}
\end{figure}

Energy gaps exhibit the unusual dependence on the concentration and distribution of O atoms, as listed in Table 1. $E_g$'s are sensitive to the change of distribution. Although their values have the non-monotonous dependence on the O-concentration, they could be divided into three categories. The GO systems always have a finite gap for the O-concentration higher than 25\%, mainly owing to the sufficiently strong O-O and C-O bonds. In the concentration range of ~25-3\%, $E_g$'s decline quickly and fluctuate widely. Such small or vanishing gaps are related to the reformed distorted Dirac  cones (Figs. 2(d)-2(g)).
Energy gaps become zero thoroughly for the low concentration of $<$3\%, being induced by the fully reformed Dirac cones without spacing (Figs. 2(h)-2(k)). In addition, the tiny and zero energy gaps are identified from the low-energy three-dimensional bands shown in the insets, in which the blue and green regions are the projections on the xz and yz planes, respectively. By adjusting the O-coverages, the band gap of graphene can be significantly modified which provides highly potential applications in nanoelectronic devices.

The charge density ($\rho$) and charge density difference ($\Delta(\rho)$) provide useful information about the chemical bondings in Figs. 3(a) and 3(b). The former clearly illustrates the terminated $\pi$ bondings and the reformed $\pi$ bondings (black rectangles of Fig. 3(a)). The later is created by substracting the charge density of isolated C and O atoms from that of GO system. The orbitals of O have strong hybridizations with those of passivated C, as seen from the green region enclosed by the dashed black rectangle (Fig. 3(b)). This region lies between them and bents toward O atoms. Such strong C-O bonds lead to the termination of the complete $\pi$ bondings between parallel $2p_z$ orbitals of C atoms causing the Dirac-cone structure near $E_F$ in pristine graphene. Between two non-passivated C atoms of GO, ($\Delta(\rho)$) shows a strong $\sigma$ bonding indicated by the enclosed pink rectangle, being slightly weakened after the formation of C-O bond.

\begin{figure}[htb]
\centering\includegraphics[width=10cm]{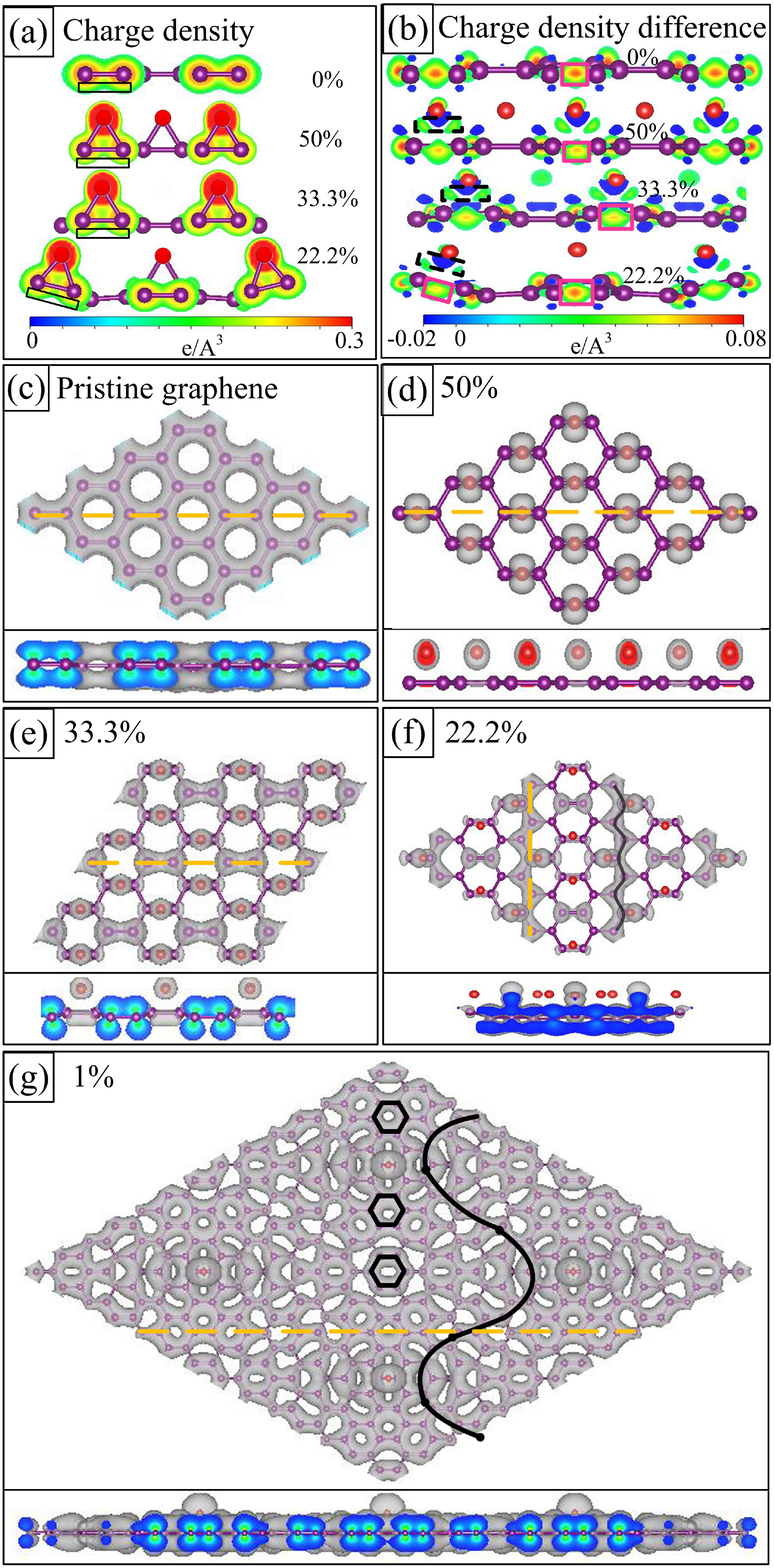}
\caption{The charge density $\rho$ (a), charge density difference $\Delta\rho$ (b), and band-decomposed chagre density for: (c) pristine graphene, (d) 50\%, (e) 33.3\%, (f) 22.2\%, and (g) 1\% O-concentrations.}
\end{figure}

The concentration-dependent energy gaps could be further understood from the 2$p_z$-orbital bondings of C atoms. The band-decomposed charge density distribution is associated with the electronic states nearest to $E_F$, which can illustrate the relation between energy gap and $\pi$ bondings. There exists five types of $\pi$ bondings as O-concentration decreases from the full to vanishing adsorptions: the complete termination (50\% system, Fig. 3.(d)), the partial suppression (33.3\% system, Fig. 3(e)), the 1D bonding (22.2\% system, Fig. 3(f)), the deformed planar bonding (sys1\% system, Fig. 3(g)), and the well-behaved one (pristine graphene, Fig. 3(c)). The competition between the weakened O-O bonds and the gradually recovered $\pi$ bondings results in the reduced band gap. The reformation of $\pi$ bondings along one direction or hexagonal rings can lead to the distorted Dirac cone recover this structure. The relation between the $\pi$-bonds and energy gap of GO is really worth mentioning, since it has not been reported elsewhere before. 

\begin{figure}[htb]
\centering\includegraphics[width=10cm]{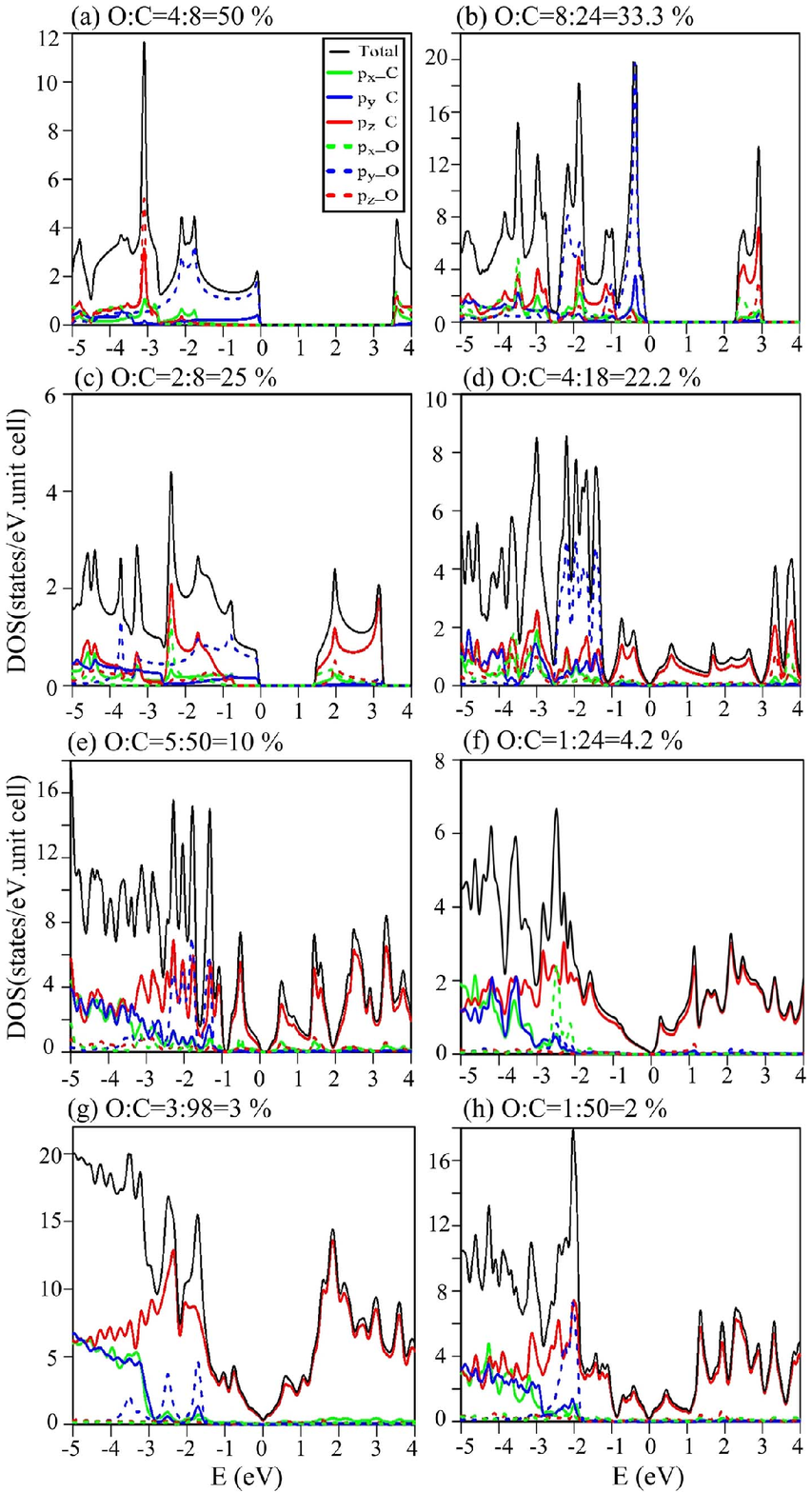}
\caption{Orbital-projected DOS of GO with various concentrations: (a) 50\%, (b) 33.3\%, (c) 25\%, (d) 22.2\%, (e) 10\%, (f) 4.2\%, (g) 3\%, and (h) 2\%.}
\end{figure}

The main characteristics of band structures are directly reflected in DOS. The orbital-projected DOS, as shown in Fig. 5, is useful in understanding the orbital contributions and hybridizations in chemical bonds. The low-energy DOS is dramatically altered after oxygen adsorption. For pristine graphene and low O-concentration (lower than 3\%; Figs. 5(g), and 5(h)), the strong peaks caused by $\pi$ and $\pi^*$ bands due to 2$p_z-$2$p_z$ bondings between C atoms will dominate within the range of $|E^{c,v}|\le\,$2 eV, indicating zero gap semiconductor behavior. However, for high O-concentration, the $\pi$- and $\pi^*$-peaks are absent as a result of the strong C-O bond (Figs. 5(a) and 5(f)). Instead, there are several O-dominated prominent structures in a wide range of -2.5 eV$\,\le\,E^v\le\,$0 eV. As the O-concentration declines, the $\pi$- and $\pi^*$-peaks are gradually reformed at low-energy, while the O-dominated structures present the narrower range and the large red shift (Figs. 5(c)-(f)). For middle-energy (-4 eV$\,\le\,E^v\le\,$-2 eV), DOS grows quickly and exhibits prominent peaks due to the C-O bonds. This means the strong hybridizations among the (2p$_x$, 2$p_z$) orbitals of O and the (2p$_x$, 2p$_y$, 2$p_z$) orbitals of passivated C atoms.

Recently, the ARPES and STS have served as efficient methods to examine the main characteristics in band structures and DOS, respectively. The former has been used to identify the Dirac-cone structure of graphene grown on SiC \cite{ohta2008morphology}, and observed the opening of band gap for graphene on Ir(111) through oxidation \cite{schulte2013bandgap}. As expected, the feature-rich energy bands of GO, including the absence and presence of the distorted Dirac cone, and the concentration-dependent energy gaps, can be verified by ARPES. The later has investigated the adatom-induced features in monolayer graphene \cite{gyamfi2011fe,tapaszto2008tuning}, and detected the band gap of exfoliated oxidized graphene sheets \cite{pandey2008scanning}. The features in DOS, mainly the finite or vanished energy gaps, the $\pi$- and $\pi^*$-peaks, can be further investigated with STS to examine whether $\pi$ bondings exist. The comparisons between theoretical predictions and experimental measurements can comprehend the relation between energy gap and $\pi$ bondings.

\vskip 0.6 truecm
\par\noindent
{\bf 4. Conclusion }
\vskip 0.3 truecm

In conclusions, the geometric structures and electronic properties of GO not only depend on the O-concentration but also the O-distribution. The low-lying electronic structures are dominated by the critical hybridizations in C-C, O-O, and C-O bonds. The band structures and band-decomposed charge densities indicate the relation between the recovered $\pi$ bondings and the energy gap of GO. Five types of $\pi$ bondings are revealed during the variation from the full to vanishing oxygen adsorptions, including the complete termination, the partial suppression, the 1D bonding, the deformed planar bonding, and the well-behaved one. They are responsible for the O-dependent energy gaps, in which the first two types cause larger band gaps for concentration of $>$25\%, the last two types lead to zero gaps for concentration of $<$3\%, and the third one relates to small or vanishing gaps for intermediate concentrations. Such energy gaps can be examined by the experimental measurements of ARPES on electronic structures and STS on DOS. The feature-rich geometric structures and electronic properties, especially for the diverse energy gaps may be significantly important for the applications in nanoelectronic devices.\\

\par\noindent {\bf Acknowledgments}

This work was supported by the National Center for Theoretical Sciences (South) and the Nation Science Council of Taiwan, under the grant No. NSC 102-2112-M-006-007-MY3. 

\newpage
\renewcommand{\baselinestretch}{0.2}

\newpage \centerline {\Large \textbf {FIGURE CAPTIONS}}

\vskip0.5 truecm 

Fig. 1. Geometric structures of GO with various unit cells and concentrations: (a) 8 C atoms per cell, (b) 18 C atoms per cell, (c) 24 C atoms per cell, (d) 50 C atoms per cell, and (e) 98 C atoms per cell. Oxygen distributions are denoted by numbers, corresponding to different concentrations listed in table 1.

Fig. 2. Band structures of GO with various concentrations: (a) 50\%, (b) 33.3\%, (c) 25\%, (d) 22.2\%, (e) 12.5\%, (f) 10\%, (g) 4.2\%, (h) 3\%, (i) 2\%, and (k) 1\%. Superscripts $c$ and $v$ correspond to the conduction and valence bands, respectively. The red and blue circles, respectively, correspond to the contributions of passivated C atoms and O atoms, in which the dominance is proportional to the radius of circle. Also shown in the inset of (a) is the first Brillouin zone.

Fig. 3. The charge density $\rho$ (a), charge density difference $\Delta\rho$ (b), and band-decomposed chagre density for: (c) pristine graphene, (d) 50\%, (e) 33.3\%, (f) 22.2\%, and (g) 1\% O-concentrations.

Fig. 4. Orbital-projected DOS of GO with various concentrations: (a) 50\%, (b) 33.3\%, (c) 25\%, (d) 22.2\%, (e) 10\%, (f) 4.2\%, (g) 3\%, and (h) 2\%.

\end{document}